\documentclass[fleqn,usenatbib]{mnras}

\usepackage{newtxtext,newtxmath}

\usepackage[T1]{fontenc}
\usepackage{ae,aecompl}

\usepackage{lipsum}
\usepackage{graphicx}	
\usepackage{amsmath}	
\usepackage{psfig}
\usepackage{xspace}
\usepackage{longtable}
\usepackage{multirow}
\usepackage{natbib}
\graphicspath{{figs/}}
\usepackage{gensymb}
\usepackage{array}

\newcommand{\sref}[1]{Section~\ref{#1}}
\newcommand{\fref}[1]{Figure~\ref{#1}}
\newcommand{\fsref}[1]{Figures~\ref{#1}}
\newcommand{\tref}[1]{Table~\ref{#1}}

\def\zl{z_{\ell}}

\def\zs{z_{\rm s}}

\newcolumntype{C}[1]{>{\centering\arraybackslash}m{#1}}

\title[SuGOHI VII: Three Strongly Lensed Quasars]{Survey~of~Gravitationally~lensed~Objects~in~HSC~Imaging~(SuGOHI) -VII.~Discovery~and~Confirmation~of~Three~Strongly~Lensed~Quasars$^{\dagger}$} 
\author[A. T. Jaelani et al.]{Anton T. Jaelani$^{1,2}$\thanks{E-mail: \href{anton@phys.kindai.ac.jp}{anton@phys.kindai.ac.jp}},
Cristian E. Rusu$^{3}$,
Issha Kayo$^{4}$,
Anupreeta More$^{5,6}$,
Alessandro \newauthor Sonnenfeld$^{5,7}$,
John D. Silverman$^{5,8}$,
Malte Schramm$^{3,9}$,
Timo Anguita$^{10,11}$,
Naohisa \newauthor Inada$^{12}$, 
Daichi Kondo$^{13}$, 
Paul L. Schechter$^{14}$,
Khee-Gan Lee$^{5}$,
Masamune Oguri$^{5,15,16}$,\newauthor 
James H. H. Chan$^{17}$,
Kenneth C. Wong$^{5}$,
and Kaiki T. Inoue$^{1}$
\medskip \\ 
$^1$Department of Physics, Kindai University, 3-4-1 Kowakae, Higashi-Osaka, Osaka 577-8502, Japan\\
$^2$Astronomy Research Group and Bosscha Observatory, FMIPA, Institut Teknologi Bandung, Jl. Ganesha 10, Bandung 40132, Indonesia\\
$^3$National Astronomical Observatory of Japan, 2-21-1 Osawa, Mitaka, Tokyo 181-0015, Japan\\
$^4$Department of Liberal Arts, Tokyo University of Technology, Ota-ku, Tokyo 144-8650, Japan\\
$^5$Kavli Institute for the Physics and Mathematics of the Universe (WPI), UTIAS, The University of Tokyo, Kashiwa, Chiba 277-8583, Japan\\
$^8$Department of Astronomy, The University of Tokyo, 7-3-1 Hongo, Bunkyo-ku, Tokyo 113-0033, Japan\\
$^9$Graduate School of Science and Engineering, Saitama University, 255 Shimo-Okubo, Sakura-ku, Saitama City, Saitama 338-8570, Japan\\
$^{10}$Departamento de Ciencias Fisicas, Universidad Andres Bello, Fernandez Concha 700, Las Condes, Santiago, Chile\\
$^{11}$Millennium Institute of Astrophysics (MAS), Nuncio Monseñor Sotero Sanz 100, Providencia, Santiago, Chile\\
$^{12}$Department of Physics, Nara National College of Technology, Yamatokohriyama, Nara 639-1080, Japan\\
$^{13}$School of Computer Science, Tokyo University of Technology, 1404-1 Katakuramachi, Hachioji City, Tokyo 192-0982, Japan\\
$^{14}$MIT Kavli Institute 37-664G, 77 Massachusetts Avenue, Cambridge, MA 02139-4307, USA\\
$^{15}$Department of Physics, The University of Tokyo, 7-3-1 Hongo, Bunkyo-ku, Tokyo 113-0033, Japan\\
$^{16}$Research Center for the Early Universe, The University of Tokyo, 7-3-1 Hongo, Bunkyo-ku, Tokyo 113-0033, Japan\\
$^{17}$Laboratoire d'Astrophysique, Ecole Polytechnique F\'{e}d\'{e}rale de Lausanne (EPFL), Observatoire de Sauverny, CH-1290 Versoix, Switzerland\\
$^{\dagger}$This paper includes data gathered with the 6.5 meter Magellan Telescopes located at Las Campanas Observatory, Chile.}
\date{Accepted XXX. Received YYY; in original form ZZZ}

\pubyear{2020}

\begin{document}
\label{firstpage}
\pagerange{\pageref{firstpage}--\pageref{lastpage}}
\maketitle

\begin{abstract}
We present spectroscopic confirmation of three new two-image gravitationally lensed quasars, compiled from existing strong lens and X-ray catalogs. Images of HSC J091843.27--022007.5 show a red galaxy with two blue point sources at either side, separated by 2.26 arcsec. This system has a source and a lens redshifts $\zs=0.804$ and $\zl=0.459$, respectively, as obtained by our follow-up spectroscopic data. CXCO J100201.50$+$020330.0 shows two point sources separated by 0.85 arcsec on either side of an early-type galaxy. The follow-up spectroscopic data confirm the fainter quasar has the same redshift with the brighter quasar from the SDSS fiber spectrum at $\zs=2.016$. The deflecting foreground galaxy is a typical early-type galaxy at a redshift of $\zl=0.439$. SDSS J135944.21+012809.8 has two point sources with quasar spectra at the same redshift $\zs=1.096$, separated by 1.05 arcsec, and fits to the HSC images confirm the presence of a galaxy between these. These discoveries demonstrate the power of the Hyper Suprime-Cam Subaru Strategic Program (HSC-SSP)'s deep imaging and wide sky coverage. Combined with existing X-ray source catalogues and follow-up spectroscopy, the HSC-SSP provides us unique opportunities to find multiple-image quasars lensed by a foreground galaxy.
\end{abstract}



\section{Introduction} \label{sec:sect1}
Over the four decades since the discovery of the first gravitationally lensed quasar \citep{Walsh1979}, these systems have proven to be invaluable probes for both astrophysics and cosmology \citep[e.g.,][]{Claeskens2002,Treu2016}. Astrophysical applications range from intermediate redshifts, with the measurement of extragalactic extinction curves \citep[e.g.,][]{Falco1999,Ostman2008}, the direct detection of dark matter substructures \citep[e.g.,][]{Dalal2002,Vegetti2012}, constraints on Warm Dark Matter \citep[WDM; e.g.,][]{Inoue+15}, microlensing studies \cite{Vernardos+19}, and the calibration of the stellar mass fundamental plane \citep[e.g.,][]{Schechter2014}, to high redshifts, with elucidating the structure of quasar accretion disks \citep[e.g.,][]{Yonehara1998,Poindexter2008}, their broad line regions \citep[e.g.,][]{Sluse2012,Fian2018}, and the coevolution of supermassive black holes with their hosts galaxies \citep[e.g.,][]{Peng2006,Ding2017}. In cosmology, statistically complete samples of lensed quasars are used to set constraints on dark energy \citep[e.g.,][]{Oguri2008,Oguri2012}, and lensed quasars with measured time delays are used to infer the Hubble-Lema\^itre constant independent from other probes \citep[e.g.,][]{Suyu2010,Bonvin2017,Wong+20}. 

While these applications are broad, they all require distinct sample selection criteria, such that studies utilising lensed quasars are limited at present by the small number of $\sim220$ known systems.\footnote{An up-to-date database is maintained at \url{https://web1.ast.cam.ac.uk/ioa/research/lensedquasars/}.} Nonetheless, the numbers have increased considerably over the past decade and a half, due to the availability of large-scale, deep multi-band optical imaging surveys such as the Sloan Digital Sky Survey \citep[SDSS,][]{York2000}, and more recently the Panoramic Survey Telescope and Rapid Response
System \citep[Pan-STARRS,][]{Chambers2016}, the Dark Energy Survey \citep[DES,][]{Abbott2018}, the Hyper Suprime-Cam Subaru Strategic Program \citep[HSC-SSP,][]{Aihara+18}, and the Dark Energy Spectroscopic Instrument (DESI) Legacy Imaging Surveys \citep{Dey+19}. Various data mining techniques have been employed to isolate the intrinsically rare lensed quasars \citep{Oguri2010b} among the vast amounts of possible contaminants. These include selecting from a parent spectroscopic catalogue and applying morphological and colour criteria \citep[e.g.,][]{Oguri+06}, using machine learning techniques on both catalogue and pixel data \citep[e.g.,][]{Agnello2015}, or engaging citizen science volunteers to visually inspect large numbers of images \citep[e.g.,][]{Sonnenfeld+20}. Ultimately however, to convincingly confirm a system as a lensed quasar requires extensive follow-up spectroscopy campaigns, which is why publishing lists of candidates is common in the literature \citep[e.g.,][]{Chan+20,Rusu2019}. 

In this paper, we present the spectroscopic confirmation of three new strongly lensed quasars compiled from previous catalogs applied to the HSC-SSP footprint: HSC J091843.27--022007.5, CXCO J100201.50+020330.0, and SDSS J135944.21+012809.8 (hereafter J0918, J1002, and J1359) at (RA Dec.) = (139.6803, $-$2.3354), (150.5063, +2.0582), and (209.9342, +1.4693), respectively. Our paper is organised as follows. \sref{sec:sect2} briefly illustrates the candidate-compilation process. In \sref{sec:sect3}, we describe the imaging analysis. We present the results of the spectroscopic follow-up observations of the systems in \sref{sec:sect4}. In \sref{sec:sect5}, we present the mass modelling of the systems, and we conclude in \sref{sec:sect6}. Throughout this paper, we adopt a concordance cosmology with $\Omega_{\rm m}= 0.27,\ \Omega_{\Lambda}=0.73,\ H_0 = 73$ km s$^{-1}$ Mpc$^{-1}$. All quoted magnitudes are on the AB system, position angles are measured East of North, and uncertainties are $1\sigma$ and are assumed to be Gaussian.

\section{Lensed Quasar Candidate Compilation} \label{sec:sect2}

The lens candidates we report here are not the result of a single systematic search. Instead, we compiled candidates from three different methods employed by several of the authors: looking for close companion object in the Chandra catalog of X-ray selected objects \citep{Evans+10}, carried out by co-author Anupreeta More, re-inspection of the Sloan Digital Sky Survey Quasar Lens Search \citep[SQLS, e.g.,][]{Oguri+06,Inada+12} catalog, conducted by co-author Issha Kayo, and further examination of the Survey of Gravitationally-lensed Objects in HSC Imaging \citep[SuGOHI,][]{Sonnenfeld+18,Sonnenfeld+19,Sonnenfeld+20,Wong+18,Chan+20,Jaelani+20}. 

As preliminary selection, we checked 5 candidates of SQLS on HSC footprint to see whether we can clearly observe an extended object (galaxy) between two point sources with quite similar colours. J1359 was the only promising candidate for spectroscopic follow-up. We selected 3 of 15 candidates from the SuGOHI lenses \citep[e.g.,][]{Chan+20,Jaelani+20,Sonnenfeld+20} based on their brightness and with suitable coordinates for follow-up.

In total, we selected 5 candidates, two from the first two methods each, and three from SuGOHI, for conducting follow-up spectroscopic observations. Details of each candidate are described in the next sections. 

\begin{figure*}
\begin{center}
\includegraphics[width=0.85\textwidth]{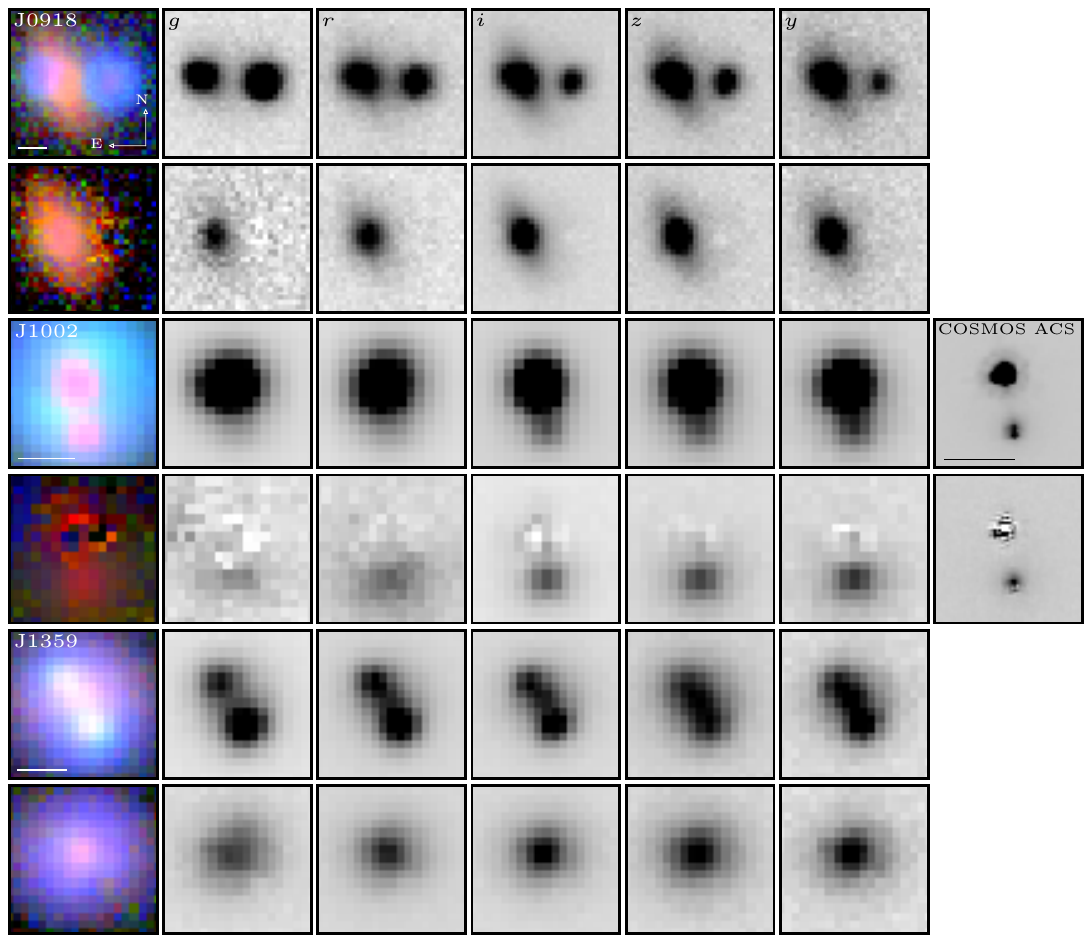}
\caption{\label{fig:figure1}HSC $gri$ composite and multiband images of the three lensed quasars. Additionally, the COSMOS ACS F814W image for J1002 is shown. The photometric bands are denoted in the upper left corner. Corresponding point source-substracted images based on the \texttt{GALFIT} best-fit in each band are shown below the data. The size of each image of J0918, J1002, and J1359 are 5.54, 2.87, and 3.19 arcsec, respectively. North is up and East is left. Scale bars of 1 arcsec are displayed in the bottom left corner.}
\end{center}
\end{figure*}

\begin{table*}
\centering
\caption{\label{tab:table1}Properties of three confirmed lensed quasars. The redshifts of lens and source are obtained from our follow-up spectra. The relative astrometry in $i$ band in arcsec, and the magnitudes of the brighter (A), the fainter (B) quasar image and the lens galaxy (G) are obtained by fitting the HSC images using \texttt{GALFIT}. The HSC image quality, in terms of the "seeing" FWHM in arcsec, is also quoted. A tentative result $\zl=0.35$ for J1359.}
\begin{tabular}{ccccccccc}
\hline
System $[\zl,\zs]$ & & $\Delta$RA & $\Delta$Dec. & $g$ & $r$ & $i$ & $z$ & $y$\\
\hline
\multirow{3}{*}{J0918 [0.459, 0.804]} & A & 0.00 & 0.00 & 20.57 $\pm$ 0.09 & 20.50 $\pm$ 0.09 & 20.73 $\pm$ 0.11 & 20.22 $\pm$ 0.05 & 20.46 $\pm$ 0.11 \\
 & B & $-$2.25 $\pm$ 0.02 & 0.22 $\pm$ 0.02 & 21.34 $\pm$ 0.13 & 21.24 $\pm$ 0.08 & 21.44 $\pm$ 0.13 & 20.99 $\pm$ 0.10 & 21.21 $\pm$ 0.06 \\
 & G & $-$1.67 $\pm$ 0.04 & 0.04 $\pm$ 0.03 & 22.12 $\pm$ 0.07 & 20.24 $\pm$ 0.06 & 19.36 $\pm$ 0.04 & 18.99 $\pm$ 0.07 & 18.78 $\pm$ 0.07 \\
Seeing & & & & 0.82 & 1.01 & 0.77 & 0.74 & 0.86 \\
\hline
\multirow{3}{*}{J1002 [0.439, 2.016]} & A & 0.00 & 0.00 & 18.97 $\pm$ 0.23 & 18.63 $\pm$ 0.20 & 18.40 $\pm$ 0.26 & 18.32 $\pm$ 0.18 & 18.29 $\pm$ 0.29 \\
 & B & 0.14 $\pm$ 0.03 & $-$0.84 $\pm$ 0.02 & 22.88 $\pm$ 0.17 & 21.79 $\pm$ 0.16 & 21.68 $\pm$ 0.20 & 21.52 $\pm$ 0.13 & 21.37 $\pm$ 0.13 \\
 & G & 0.13 $\pm$ 0.03 & $-$0.74 $\pm$ 0.04 & 23.84 $\pm$ 0.15 & 22.23 $\pm$ 0.14 & 21.14 $\pm$ 0.11 & 20.83 $\pm$ 0.12 & 20.71 $\pm$ 0.11 \\
Seeing & & & & 0.78 & 0.92 & 0.59 & 0.62 & 0.61 \\
\hline
\multirow{3}{*}{J1359 [ -- , 1.096]} & A & 0.00 & 0.00 & 21.03 $\pm$ 0.19 & check20.58 $\pm$ 0.12 & 20.58 $\pm$ 0.17 & 20.63 $\pm$ 0.11 & 20.43 $\pm$ 0.11 \\
 & B & $-$0.54 $\pm$ 0.02 & 0.90 $\pm$ 0.02 & 21.80 $\pm$ 0.13 & 21.33 $\pm$ 0.16 & 21.31 $\pm$ 0.18 & 21.35 $\pm$ 0.12 & 21.08 $\pm$ 0.16 \\
 & G & $-$0.23 $\pm$ 0.03 & 0.55 $\pm$ 0.04 & 21.43 $\pm$ 0.05 & 20.22 $\pm$ 0.13 & 19.75 $\pm$ 0.08 & 19.31 $\pm$ 0.10 & 19.08 $\pm$ 0.12 \\
Seeing & & & & 0.70 & 0.61 & 0.57 & 0.66 & 0.55 \\
\hline
\end{tabular}
\end{table*}

\section{Imaging} \label{sec:sect3}

In this work, we use photometric data from the HSC Wide S18A internal data release of the HSC survey \citep{Aihara+19}. The HSC Survey is an ongoing imaging survey, expected to cover about 1,400 deg$^2$ in five bands ($g, r, i, z$ and $y$) down to $r\sim 26$ with the Hyper Suprime-Cam \citep{Aihara+18,Miyazaki+18,Komiyama+18, Kawanomoto+18, Furusawa+18}, using a wide-field (1.7-degree diameter) optical camera installed on the 8.2-m Subaru Telescope. The data are processed with \texttt{hscPipe}, which is derived from the Vera C. Rubin Observatory's Legacy Survey of Space and Time pipeline \citep{Axelrod+10,Juric+17,Ivezic+08,Ivezic+19}. 

We model and analyse the multiband HSC imaging data for the lens systems using \texttt{GALFIT} \citep{Peng+02}. For each system, we measure relative positions of the fainter quasar and the lens galaxy with respect to the brighter quasar, by fitting a model composed of a point source for each quasar image and a S\'{e}rsic profile convolved with the point spread function (PSF) for the lens galaxy. The PSF models are generated using the \texttt{PSF Picker} tool \citet[see details in][]{Aihara+18b,Aihara+19} for all of the HSC bands. Magnitudes of each component, derived from the fits, are given in \tref{tab:table1}. 

In \fref{fig:figure1}, we show the HSC multiband imaging cutouts as well as the PSF-subtracted images, revealing the lens galaxies. In general, the lenses are brighter in $y$ band and the quasars in $g$ band, as expected from their red and blue colours, respectively. We show an additional image of J1002 from data archive of the Hubble Space Telescope (HST) Advanced Camera for Surveys (ACS) Wide Field Channel (WFC), part of the Cosmic Evolution Survey \citep[COSMOS,][]{Scoville+07a,Scoville+07,Koekemoer+07} with 5$\sigma$ detection limits for point sources of 27.2 AB magnitude in F814W, hereafter COSMOS ACS. 

\section{Spectroscopic Confirmation} \label{sec:sect4}

Here, we describe our follow-up spectroscopic results for each lens system. All observations were processed using standard techniques within \texttt{IRAF}, and flux-calibrated using observations of spectrophotometric standards.

\begin{figure}
\begin{center}
\includegraphics[width=\columnwidth]{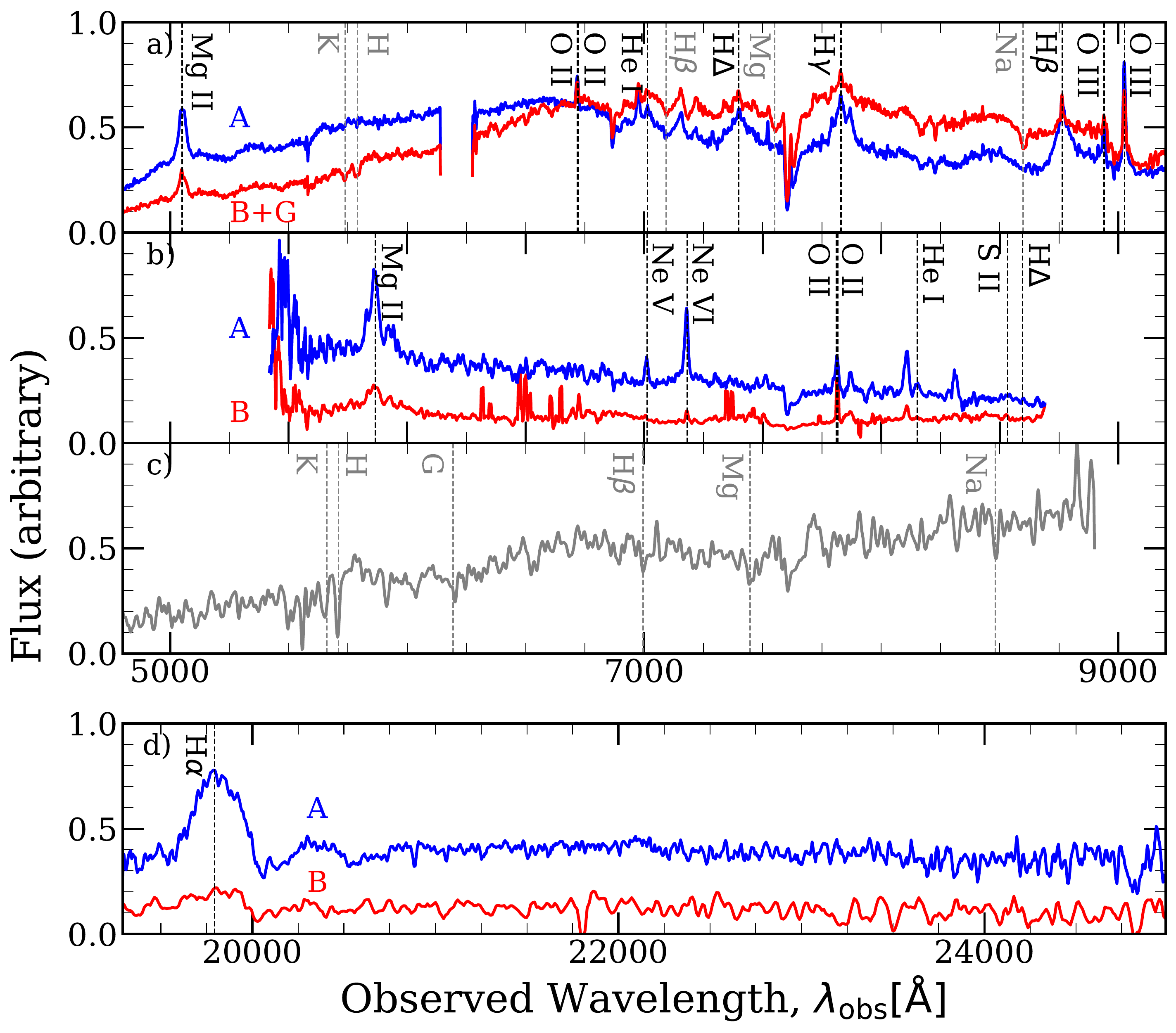}\\
\caption{\label{fig:figure2}1D spectra of J0918 (panel a), J1002 (panel c, d), and J1359 (panel b). Panel b shows only the "red part" of the LRIS spectra, with range 5500 - 8500\AA. The blue and red lines show the two components, demonstrating that they are a nearly identical quasar spectrum, and the grey line in panel c is the lens galaxy spectra. Emission and absorption lines are marked by black and grey dashed vertical lines, respectively.}
\end{center}
\end{figure}

\subsection{J0918}
J0918 system was first discovered by \citet{Sonnenfeld+20} as part of the citizen science project {\sc SpaceWarps} \citep{Marshall+16,More+16} using over 442 deg$^2$ of data from the HSC survey. An expert volunteer, Claude Cornen, identified two blue point-like feature and a putative galaxy in between. A spectrum was obtained with the Inamori Magellan Areal Camera and Spectrograph \citep[IMACS;][]{Dressler+11} on the Baade 6.5-m telescope on 2019 April 1 and 2 (Director’s Discretionary Time, PI: P. Schechter) for a total exposure time 2 $\times$ 1800 s using f/2 camera, 0.9 arcsec wide-slit, and 200 mm$^{-1}$ grism. 

The 1D spectra of the two point sources correspond to a nearly identical quasar spectrum at $\zs=0.804\pm0.001$, based on Mg {\sc ii}, O {\sc ii}, He {\sc i}, H$\delta$, H$\gamma$, H$\beta$, O {\sc iii} lines. The blended absorption lines of the lens galaxy with the fainter quasar B can be seen clearly based on Ca K, Ca H, H$\beta$, Mg b, and Na D, yielding a lens galaxy redshift of $\zl=0.459$ (see panel a in \fref{fig:figure2}). Together with the presence of a red galaxy in HSC images, the spectra confirm J0918 as a strongly lensed quasar, with image separation $ 2.26$ arcsec. 

\subsection{J1002} 

\begin{figure}
\begin{center}
\includegraphics[width=0.95\columnwidth]{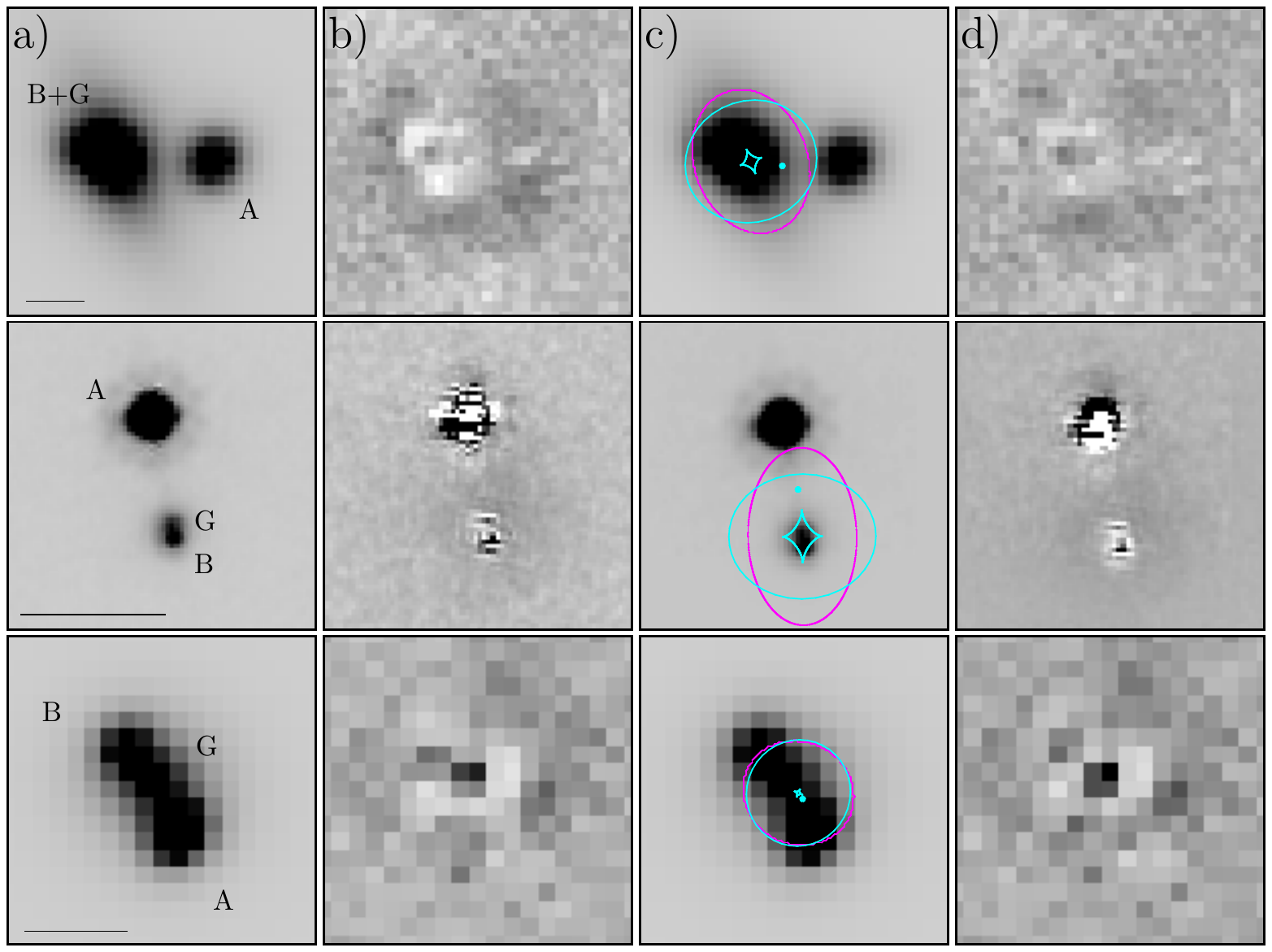}\\
\caption{\label{fig:figure3}Panel a: Model images generated from \texttt{GALFIT} using a PSF for the source images and a S\'{e}rsic model convolved with the PSF for the lens galaxy, showing the positions of the doubly-lensed quasar images (A, B) and the early-type lens galaxy (G) for each system (from top to bottom, J0918, J1002 and J1359, respectively). The black solid bar at the bottom left of the panel shows a scale of 1 arcsec. Panel b: \texttt{GALFIT} model-subtracted residual images. Panel c: Model images generated from lens mass modelling of the lens system, where the source is assumed to be a point source. 
Magenta and cyan lines indicate critical curves and caustic, respectively. The true source positions are shown by cyan dots. Panel d: The residual image obtained by subtracting the model from the data.}
\end{center}
\end{figure}

This candidate was chosen from the Chandra catalog of X-ray selected objects \citep{Evans+10}. These X-ray selected AGN sources were inspected in the optical imaging search for the presence of any close companions \citep[e.g.,][]{Liu+13}. This system was first detected in the Sloan Digital Sky Survey (SDSS) as SDSS J100201.51+020329.4 as a quasar with spectroscopic redshift of 2.016. Furthermore, this system also has CFHTLS optical and near-infrared imaging, and the COSMOS-Very Large Array (1.4 GHz) imaging in the radio shows a detection. However, the spatial resolution and sensitivity of these data are not enough to resolve this system. The HSC images of J1002, however, unambiguously reveal the presence of two sources with image separation $\approx 0.8$ arcsec. This is an excellent example of HSC images being ideal for finding small separation lenses. In the COSMOS ACS F814W imaging, we further detect emission from a third object, which is acting as the lens galaxy (see \fsref{fig:figure1} and \ref{fig:figure3}). With \texttt{GALFIT}, we measure magnitudes of A $= 18.52$, B $=22.03$, and G $=21.38$ mag. 

A near-infrared $K$-band (22,000\AA) spectrum was obtained with the Infrared Camera and Spectrograph \citep[IRCS;][]{Tokunaga+98,Kobayashi+00} at Subaru Telescope for a total exposure time 10 $\times$ 300 s using a setup of $0.45\times 18.0$ arcsec slit, 52 miliarcsec mode, $R\sim400$, and $K$-grism that covers the 19,300 - 24,800\AA~ range, assisted with the Adaptive Optics system AO188 \citep{Hayano+08,Hayano+10}, on 2012 November 27 (o12438, PI: M. Schramm). The 1D spectra of the two sources show the same broad emission line of H$\alpha$ at $\zs=2.016\pm 0.003$. The results suggest that the fainter quasar is highly likely to be at the same redshift as the brighter quasar (see panel d of \fref{fig:figure2}).

We also carried out deep optical spectroscopy to obtain the redshift of the lens galaxy. We used the Faint Object Camera and Spectrograph \citep[FOCAS;][]{Kashikawa+00} at Subaru Telescope on 2014 October 23 (o14500, PI: A. More). The observations were made with 1 arcsec long-slit mode, a 300R grating, and an SO58 order-cut filter, giving a spectral coverage of 5800-10000\AA~ with a dispersion of 1.34\AA~pixel$^{-1}$ for a total exposure time 3 $\times$ 900 s. We detect absorption lines from the lens galaxy based on Ca K, Ca H, G-band, H$\beta$, Mg b, and Na D at $\zl=0.439\pm 0.001$ (see panel c of \fref{fig:figure2}).

\subsection{J1359}
J1359 system was listed as a likely lens candidate by the Sloan Digital Sky Survey Quasar Lens Search \citep[SQLS, e.g.,][]{Oguri+06,Inada+12}, however follow-up observations had not been performed due to the small separation ($\leq1$ arcsec). We reinspected the SQLS candidate lists using HSC images and re-graded this system as a highly likely candidate. We can clearly observe an extended object (galaxy) between two point sources with quite similar colours (see \fref{fig:figure1}). The quasar was first detected in the SDSS as SDSS J135944.21+012809.8 and has a spectroscopic redshift of 1.096. This system was including as a part of program to confirm dual quasars i.e., physical pairs \citep{Silverman+20}.

We performed spectroscopic follow-up using the dual beam Low Resolution Imaging Spectrometer \citep[LRIS,][]{Oke+95,Steidel+04} at the Keck 1 telescope on 2019 January 10 as a backup target during the COSMOS Ly$\alpha$ Mapping And Tomography Observations (CLAMATO) Survey \citep[][U095, PI White]{Lee+18}. We exposed for 600s in the blue and 590s in the red using the 1 arcsec width long-slit, the 5600\AA~ dichroic, the 600 $\ell$ mm$^{-1}$ blue grism ($\lambda_{\rm blaze}=4000$\AA), and the 400 $\ell$ mm$^{-1}$red grating ($\lambda_{\rm blaze}=8500$\AA). In panel b of \fref{fig:figure2}, we can see separated spectra of the brighter and the fainter quasar, both at $\zs=1.096\pm 0.001$ from the identification of several emission lines, Mg {\sc ii}, Ne {\sc v}, Ne {\sc vi}, O {\sc ii}, He {\sc i}, S {\sc ii}, and H$\Delta$, in the red spectra. Unfortunately, we cannot extract the lens galaxy information in the blue spectra due to bad seeing. We find a tentative hint of the 4000\AA~ break feature, which corresponds to redshift $\zl\approx0.35$. This is also consistent with our rough estimation of the lens redshift, based on the HSC colours.

\begin{table}
\centering
\caption{\label{tab:table2}Best-fit model parameters with $1\sigma$ errors from MCMC: SIE velocity dispersion $\sigma_{\rm SIE}$, ellipticity $e_{\rm SIE}$, position angle $\theta_{e_{\rm SIE}}$, reduced $\chi^2$, degree of freedom $N_{\rm DOF}$, total source magnification $\mu_{\rm tot}$, time delay between the two lensed images $\Delta t_{\rm AB}$, Einstein radius $\theta_{\rm Ein}$, and enclosed mass $M_{\rm Ein}$.}
\begin{tabular}{lccc}
\hline
Lens Parameter & J0918 & J1002 & J1359 \\
\hline
$\sigma_{\rm SIE}$ (km~s$^{-1}$) & $321\pm 6$ & $155\pm 3$ & $172\pm 5$\\
$e_{\rm SIE}$ & 0.23$\pm$0.04 & 0.39$\pm$0.03 & 0.10$\pm$0.03 \\
$\theta_{e_{\rm SIE}}$ (deg) & $21\pm5$ & $1\pm4$ & $66\pm5$\\
$\chi_{\rm red}^2$ & 1.16 & 2.27 & 1.12\\
$N_{\rm DOF}$ & 1284 & 2847 & 789\\
$\mu_{\rm tot}$ & $4.1\pm1.9$ & $4.9\pm2.1$ & $10.8\pm3.0$\\
$\Delta t_{\rm AB}$, days & $-141.1\pm5.7$ & $-18.7\pm4.6$ & $-5.7\pm1.7$ \\
$\theta_{\rm Ein}$, arcsec & $1.28\pm 0.06$ & $0.61\pm 0.03$ & $0.55\pm 0.04$ \\
$M_{\rm Ein}$, $10^{11}h^{-1}M_{\odot}$ & $6.50\pm0.50$ & $0.62\pm0.05$ & $0.74\pm0.09$ \\
\hline
\end{tabular}
\end{table}

\section{Lensing Properties}\label{sec:sect5}
We used the publicly available lens modelling software \texttt{glafic} \citep{Oguri+10} to model the lensing configuration of the three systems. In order to explore the parameter posteriors, we ran a custom code that combines \texttt{glafic} and the \texttt{emcee} \citep{Foreman+13} to perform Markov Chain Monte Carlo (MCMC) sampling. With the exception of J1002, where we fitted the COSMOS data, we used the flux information in image pixels (within 7.56, 2.25, and 5.88 arcsec of J0918, J1002, and J1359, respectively) of the lens galaxies and the quasar images in the HSC $i$ band image, which has better seeing, as constraints for the lens mass model.

Light from the lens galaxy and the unresolved source quasars is modelled with a seven-parameter elliptical S\'{e}rsic profile and a three-parameter PSF profile, respectively. We used the best-fitting parameters from \texttt{GALFIT} with prior knowledge of the position, the ellipticity, and position angle from $i$ band as initial parameters input to \texttt{glafic} to model the lens galaxy G. For the source, we let every parameter of the PSF profile vary: magnitude and as well as positions $\rm x_{s}$, $\rm y_{s}$. We consider a singular isothermal ellipsoid (SIE), with surface density given by $\kappa=\theta_{\rm Ein}/\left(2\sqrt{\tilde{x}^2+\tilde{y}^2/(1-e)^2}\right)$ where $(\tilde{x}, \tilde{y})$ are the coordinates relative to the lens center along the semi-major and semi-minor axes of the elliptical mass distribution, for the lens mass model. The mass distribution is optimized in the image plane by minimizing $\chi^2$ the flux information between the observed and lens model image. The inferred angular Einstein radius and enclosed mass is defined by

\begin{equation}\label{eq:Einseq}
\theta_{\rm Ein}=4\pi\left( \frac{\sigma_{\rm SIE}}{c}\right)^2\frac{D_{\rm \ell s}}{D_{\rm s}}
\end{equation}
and
\begin{equation}\label{eq:Meinseq}
M_{\rm Ein}=\pi\Sigma_{\rm cr}(\theta_{\rm Ein}D_{\ell})^2
\end{equation}
with $\Sigma_{\rm cr}=c^2D_{\rm s}/(4\pi G D_{\rm \ell}D_{\rm \ell s})$ in terms of angular diameter distance to the source $D_{\rm s}$, to the lens $D_{\rm \ell}$, between the observer and the source $D_{\rm \ell s}$, with gravitational constant $G$, and the speed of light $c$. The results of lens modelling are summarised in \tref{tab:table2} and \fref{fig:figure3} for $i$ band. Note for J1359 we used an estimated lens redshift $\zl=0.35$ for the lens mass model.

\section{Conclusions} \label{sec:sect6}
We have presented the spectroscopic confirmation of three new doubly lensed quasars, compiled from distinct catalogs. The HSC images of J0918 and J1359 reveal two quasar images on either side of lensing source, separated by 2.26 and 1.05 arcsec, respectively, and a close companion for J1002. COSMOS ACS image shows two resolved quasar images and a putative lens galaxy for J1002. Spectra confirm that each system is a lensing system with $\zs=0.804$, 2.016, and $\zl=0.459$, 0.439, for J0918 and J1002, respectively. J1359 has two quasar spectra at $\zs=1.096$ with similar features and the lens spectrum cannot be extracted due to poor seeing during observation. 

We fit SIE lens models to the lenses and infer the Einstein radius and enclosed mass. J1002 and J1359 have velocity dispersion $\sigma_{\rm SIE}=155$ and 172 km s$^{-1}$ corresponding to enclosed mass within $\theta_{\rm Ein}$ of $M_{\rm Ein}=0.62\times10^{11}$ and $0.74\times10^{11}$ $h^{-1}M_{\odot}$, respectively. Hence, these systems provide the opportunity to study galaxies from the lower end of the mass function compared to the typical galaxies that have been studied with gravitational lensing so far.

The discovery of these lenses shows us the importance of reanalysis or re-inspection of previous lens search catalogues, by comparing to ongoing or upcoming imaging surveys of superior quality. We have shown that the unique combination of deep, wide imaging, high pixel resolution and good seeing conditions of the HSC data make possible to discover small separation lenses as part of the ongoing SuGOHI search. 

\section*{Acknowledgements}
This work was supported by JSPS KAKENHI Grant Number JP17H02868 (ATJ, KTI), JP15H05892, JP18K03693 (MO), JP15H05896, JP20K04016 (IK), and JP18H05868 (KGL). TA acknowledges support from Proyecto Fondecyt N1190335 and the Ministry for the Economy, Development, and Tourism's Programa Inicativa Cient\'{i}fica Milenio through grant IC 12009. JHHC acknowledges support from the Swiss National Science Foundation (SNSF). We would like to thank Sherry Suyu and Chien-Hsiu Lee for useful comments and suggestions. We also thank Martin White for letting us use the the spectroscopic data of J1359.

The Hyper Suprime-Cam (HSC) collaboration includes the astronomical communities of Japan and Taiwan, and Princeton University. The HSC instrumentation and software were developed by the National Astronomical Observatory of Japan (NAOJ), the Kavli Institute for the Physics and Mathematics of the Universe (Kavli IPMU), the University of Tokyo, the High Energy Accelerator Research Organization (KEK), the Academia Sinica Institute for Astronomy and Astrophysics in Taiwan (ASIAA), and Princeton University. Funding was contributed by the FIRST program from the Japanese Cabinet Office, the Ministry of Education, Culture, Sports, Science and Technology (MEXT), the Japan Society for the Promotion of Science (JSPS), Japan Science and Technology Agency (JST), the Toray Science Foundation, NAOJ, Kavli IPMU, KEK, ASIAA, and Princeton University.

This paper makes use of software developed for the Large Synoptic Survey Telescope. We thank the LSST Project for making their code available as free software at \href{http://dm.lsst.org}{http://dm.lsst.org}.

This paper is based [in part] on data collected at the Subaru Telescope and retrieved from the HSC data archive system, which is operated by Subaru Telescope and Astronomy Data Center (ADC) at NAOJ. Data analysis was in part carried out with the cooperation of Center for Computational Astrophysics (CfCA), NAOJ.

The Pan-STARRS1 Surveys (PS1) and the PS1 public science archive have been made possible through contributions by the Institute for Astronomy, the University of Hawaii, the Pan-STARRS Project Office, the Max Planck Society and its participating institutes, the Max Planck Institute for Astronomy, Heidelberg, and the Max Planck Institute for Extraterrestrial Physics, Garching, The Johns Hopkins University, Durham University, the University of Edinburgh, the Queen’s University Belfast, the Harvard-Smithsonian Center for Astrophysics, the Las Cumbres Observatory Global Telescope Network Incorporated, the National Central University of Taiwan, the Space Telescope Science Institute, the National Aeronautics and Space Administration under grant No. NNX08AR22G issued through the Planetary Science Division of the NASA Science Mission Directorate, the National Science Foundation grant No. AST-1238877, the University of Maryland, Eotvos Lorand University (ELTE), the Los Alamos National Laboratory, and the Gordon and Betty Moore Foundation.

\section*{Data Availability}

The basic information of the three lensed quasars can be browsed or downloaded from the SuGOHI online database \url{http://www-utap.phys.s.u-tokyo.ac.jp/~oguri/sugohi/} and imaging data from the HSC-SSP Public Data Release 2 \url{https://hsc-release.mtk.nao.ac.jp/doc/}. The spectroscopic data will be shared upon reasonable request to the corresponding author.

\bibliographystyle{mnras}
\bibliography{references_papers}

\appendix
\section{Rejected candidates}\label{app:rejected}

\begin{figure}
\begin{center}
\includegraphics[width=\columnwidth]{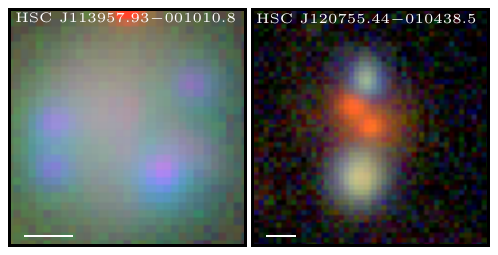}\\
\caption{\label{fig:app} HSC $gri$ composite images of the the two rejected candidates. Scale bars of 1 arcsec are displayed in the bottom left corner.}
\end{center}
\end{figure}

By using the same instrument setup for observing J0918 spectra, we observed two other candidates HSC J113957.93$-$001010.8 and HSC J120755.44$-$010438.5, for a total exposure time 1800 s per system (see \fref{fig:app}), which we convincingly reject. The first candidate was made up of four blue components in a quad-like configuration, with odd flux ratios. The close pair should have been brighter and the quadrilateral made by the four images has two opposite sides that are twice as big as the other pair of opposite sides. The spectrum of the two brightest components showed ionized knots with H$\beta$, O {\sc iii} and H$\alpha$ which indicate a bunch of star forming regions likely associated to a galaxy at $z\sim0.05$. The spectrum of second candidate likewise showed narrow H$\beta$ and O {\sc iii} at $z=0.37$. 

\bsp	
\label{lastpage}
\end{document}